\documentclass[aps,preprintnumbers,superscriptaddress,showpacs]{revtex4}
\usepackage{epsfig}
\usepackage{psfrag}
\usepackage{amsfonts}
\usepackage{graphicx}
\usepackage{dcolumn}
\usepackage{bm}

\begin{document}

\title{Holographic Schwinger effect in a confining D3-brane background with chemical potential}

\author{Zi-qiang Zhang}
\email{zhangzq@cug.edu.cn} \affiliation{School of mathematical and
physics, China University of Geosciences(Wuhan), Wuhan 430074,
China}

\author{De-fu Hou}
\email{houdf@mail.ccnu.edu.cn} \affiliation{Key Laboratory of
Quark and Lepton Physics (MOE), Central China Normal University,
Wuhan 430079,China}

\author{Yan Wu}
\email{yan.wu@cug.edu.cn} \affiliation{School of mathematical and
physics, China University of Geosciences(Wuhan), Wuhan 430074,
China}

\author{Gang Chen}
\email{chengang1@cug.edu.cn} \affiliation{School of mathematical
and physics, China University of Geosciences(Wuhan), Wuhan 430074,
China}

\begin{abstract}
Using the AdS/CFT correspondence, we investigate the Schwinger
effect in a confining D3-brane background with chemical potential.
The potential between a test particle pair on the D3-brane in an
external electric field is obtained. The critical field $E_c$ in
this case is calculated. Also, we apply numerical method to
evaluate the production rate for various cases. The results imply
that the presence of chemical potential tends to suppress the pair
production effect.
\end{abstract}

\maketitle
\section{Introduction}
Schwinger effect is known as the pair production in an external
electric field in QED \cite{JS}. The virtual electron-position
pairs can become real particles when a strong electric-field is
applied. The production rate $\Gamma$ has been calculated in the
weak-coupling and weak-field approximation long time ago. Later,
it is generalized to arbitrary-coupling and weak-field case in
\cite{IK}
\begin{equation}
\Gamma\propto e^{\frac{-\pi m^2}{eE}+\frac{e^2}{4}},
\end{equation}
where m and e are the mass and charge of the created particles
respectively, E is the external electric field. This
non-perturbative effect can be explained as a tunneling process,
and is not restricted to QED but usual for QFTs coupled to a U(1)
gauge field.

AdS/CFT, namely the duality between the type IIB superstring
theory formulated on AdS$_5\times S^5$ and $\mathcal N=4$ SYM in
four dimensions, can realize a system that coupled with a U(1)
gauge field \cite{Maldacena:1997re,Gubser:1998bc,MadalcenaReview}.
Therefore, it is very interesting to consider the Schwinger effect
in a holographic setup. After the works \cite{AS,GW} in which the
creation rate of the quark pair in $\mathcal N=4$ SYM theory was
obtained firstly, there are many attempts to addressing Schwinger
effect in this direction. For instance, the universal aspects of
this effect in the general backgrounds are investigated in
\cite{YS}. The pair production in confining background is studied
in \cite{YS1,DK}. The potential barrier for the pair creation is
analyzed in \cite{YS2}. The pair production in the conductivity of
a system of flavor and color branes is analyzed in \cite{SC}. The
Schwinger effect with constant electric and magnetic fields is
investigated in \cite{SB,YS3}. The non-relativistic Schwinger
effect is discussed in \cite{KB}. The Gauss-Bonnet corrections to
this effect is studied in \cite{SZ}. This effect is also studied
with some AdS/QCD models \cite{KH,JS1}. Other important results
can be found for example in \cite{JA,KH1,DD,WF,MG,XW}. For a
recent review on this topic, see \cite{DK1}.

As we know, a finite temperature of the gauge theory is on the
gravity side of the duality represented by a black-hole horizon in
the bulk located at $r_t$ in the fifth coordinate and extended in
the other four space-time directions. In addition, a chemical
potential of the gauge theory can be obtained in this setup by
giving the black hole an electric charge. Since the Schwinger
effect in a confining D3-brane background has been discussed in
\cite{YS1}, it is interesting to consider the influence of the
chemical potential on Schwinger effect in this case. In this
paper, we will add the chemical potential to the confining
D3-brane background. We would like to see how chemical potential
affects the Schwinger effect. It is the motivation of the present
work.

The organization of this paper is as follows. In the next section,
the background with chemical potential is briefly introduced. In
section III, we perform the potential analysis in this background.
Then we investigate the production rate in section IV. The last
part is devoted to conclusion and discussion.

\section{Background with chemical potential}
In the framework of the AdS/CFT duality, $\mathcal N=4$ SYM theory
with a chemical potential is obtained by making the black hole in
the holographic dimension be charged. The corresponding metric is
an asymptotically $AdS_5$ space with a Reissner-Nordstrom black
hole. Its metric can be written as \cite{CE}
\begin{eqnarray}
ds^2
=\frac{r^2}{L^2}[-(dx^0)^2+(dx^1)^2+(dx^2)^2+f(r)(dx^3)^2]+\frac{L^2}{r^2}f(r)^{-1}dr^2+L^2d\Omega_5^2,\label{metric}
\end{eqnarray}
with
\begin{equation}
f(r)=1-(1+Q^2)(\frac{r_t}{r})^4+Q^2(\frac{r_t}{r})^6,
\end{equation}
where $L$ is the AdS space radius, the string tension
$\frac{1}{2\pi\alpha^\prime}$ relates the 't Hooft coupling
constant by $\frac{L^2}{\alpha^\prime}=\sqrt{\lambda}$, r denotes
the radial coordinate of the black brane geometry. $r_t$ is the
inverse compactification radius in the $x^3$-direction. As $r_t$
grows, the radius becomes shrink.

The chemical potential $\mu$ is given by
\begin{equation}
\mu=\frac{\sqrt{3}Qr_t}{L^2},
\end{equation}
where the charge Q of the black hole is in the range $0\leq
Q\leq\sqrt{2}$.

Note that, when $Q=0$ (zero chemical potential), the usual
confining D3-brane background is reproduced. In addition, when
$r_t=0$ (zero temperature), the usual $AdS_5\times S^5$ background
is reproduced.

We should point out that the chemical potential implemented in
this way is not the quark (or baryon) chemical potential of QCD
but a chemical potential that refers to the R-charge of $\mathcal
N=4$ SYM theory. However, we can apply it as a simple way of
introducing finite density effects into the system in the present
work. More discussions about the AdS-RN in the context of
holographic set-ups can be found in \cite{MC,AC}.

\section{potential analysis}

\subsection{potential  research}
We now perform the potential analysis for confining D3-brane
background with chemical potential using the metric
Eq.(\ref{metric}). The classical string action can reduce to the
Nambu-Goto action
\begin{equation}
S=T_F\int d^2\sigma\sqrt{g}=T_F\int d\tau d\sigma\mathcal L,
\qquad T_F=\frac{1}{2\pi\alpha^\prime},
\end{equation}
where $g$ is the determinant of the induced metric on the string
world sheet embedded in the target space, i.e.
\begin{equation}
g_{\alpha\beta}=g_{\mu\nu}\frac{\partial
X^\mu}{\partial\sigma^\alpha} \frac{\partial
X^\nu}{\partial\sigma^\beta},
\end{equation}
where $X^\mu$ and $g_{\mu\nu}$ are the target space coordinates
and the metric, and $\sigma^\alpha$ with $\alpha=0,1$ parameterize
the world sheet.

By using the static gauge
\begin{equation}
x^0=\tau, \qquad x^1=\sigma,\qquad r=r(\sigma).
\end{equation}

We can obtain the induced metric $g_{\alpha\beta}$
\begin{equation} g_{00}=\frac{r^2}{L^2}, \qquad
g_{01}= g_{10}=0,\qquad
g_{11}=\frac{r^2}{L^2}+\frac{L^2}{r^2[1-(1+Q^2)(\frac{r_t}{r})^4+Q^2(\frac{r_t}{r})^6]}(\frac{\partial
r}{\partial\sigma})^2.
\end{equation}

Then the lagrangian density becomes
\begin{equation}
\mathcal
L=\sqrt{\frac{1}{1-(1+Q^2)(\frac{r_t}{r})^4+Q^2(\frac{r_t}{r})^6}\dot{r}^2+\frac{r^4}{L^4}},\label{L}
\end{equation}
with $\dot{r}=dr/d\sigma$.

Now that $\mathcal L$ does not depend on $\sigma$ explicitly, we
have the conserved quantity,
\begin{equation}
\mathcal L-\frac{\partial\mathcal L}{\partial\dot{r}}\dot{r}.
\end{equation}

The boundary condition at $\sigma=0$ is
\begin{equation}
\dot{r}=0,\qquad  r=r_c\qquad (r_t<r_c<r_0),
\end{equation}
which leads to
\begin{equation}
\frac{r^4}{\sqrt{\frac{1}{1-(1+Q^2)(\frac{r_t}{r})^4+Q^2(\frac{r_t}{r})^6}\dot{r}^2+\frac{r^4}{L^4}}}=C=r_c^2L^2,
\end{equation}
then a differential equation is derived,
\begin{equation}
\dot{r}=\frac{dr}{d\sigma}=\frac{r^2\sqrt{r^4-r_c^4}}{r_c^2L^2}\sqrt{1-(1+Q^2)(\frac{r_t}{r})^4+Q^2(\frac{r_t}{r})^6}\label{dotr}.
\end{equation}

By integrating Eq.(\ref{dotr}) the separate length $x$ of the test
particle pair on the D3 brane becomes
\begin{equation}
x=\frac{2L^2}{r_0a}\int_1^{1/a}\frac{dy}{y^2\sqrt{(y^4-1)[1-(1+Q^2)(\frac{b}{ay})^4+Q^2(\frac{b}{ay})^6]}}\label{xx},
\end{equation}
where the following dimensionless parameters have been introduced,
\begin{equation}
y\equiv\frac{r}{r_c},\qquad a\equiv\frac{r_c}{r_0},\qquad
b\equiv\frac{r_t}{r_0}.
\end{equation}

Plugging Eq.(\ref{dotr}) into Eq.(\ref{L}), one obtains the
classical action. Then the sum of potential energy(PE) and static
energy(SE) of string is obtained as
\begin{equation}
V_{PE+SE}=2T_Fr_0a\int_1^{1/a}\frac{y^2dy}{\sqrt{(y^4-1)[1-(1+Q^2)(\frac{b}{ay})^4+Q^2(\frac{b}{ay})^6]}}.\label{en}
\end{equation}

\subsection{the critical field}
To ensure that the potential analysis is right and consistent with
the DBI result, we should pause here to gain the critical field by
DBI action at hand.

The DBI action is
\begin{equation}
S_{DBI}=-T_{D3}\int
d^4x\sqrt{-det(G_{\mu\nu}+\mathcal{F}_{\mu\nu})}\label{dbi},
\end{equation}
where $T_{D3}$ is the D3-brane tension
\begin{equation}
T_{D3}=\frac{1}{g_s(2\pi)^3\alpha^{\prime^2}}
\end{equation}

In terms of the metric Eq.(\ref{metric}), the induced metric
$G_{\mu\nu}$ reads
\begin{equation}
G_{00}=-\frac{r^2}{L^2}, \qquad G_{11}=\frac{r^2}{L^2},\qquad
G_{22}=\frac{r^2}{L^2},\qquad
G_{33}=\frac{r^2}{L^2}[1-(1+Q^2)(\frac{r_t}{r})^4+Q^2(\frac{r_t}{r})^6]=\frac{r^2}{L^2}f(r).
\end{equation}

After considering the $\mathcal{F}_{\mu\nu}$ term, which can be
written as $\mathcal{F}_{\mu\nu}=2\pi\alpha^\prime F_{\mu\nu}$
\cite{BZ}, we find
\begin{equation}
G_{\mu\nu}+\mathcal{F}_{\mu\nu}=\left(
\begin{array}{cccc}
 -\frac{r^2}{L^2} & 2\pi\alpha^\prime E_1 & 2\pi\alpha^\prime E_2 & 2\pi\alpha^\prime E_3\\
 -2\pi\alpha^\prime E_1 & \frac{r^2}{L^2} & 0 & 0 \\
 -2\pi\alpha^\prime E_2 & 0 & \frac{r^2}{L^2} & 0\\
 -2\pi\alpha^\prime E_3 & 0 & 0 & \frac{r^2}{L^2}f(r)
\end{array}
\right),
\end{equation}
which leads to
\begin{equation}
det(G_{\mu\nu}+\mathcal{F}_{\mu\nu})=-\frac{r^4}{L^4}f(r)[\frac{r^4}{L^4}-(2\pi
\alpha')^2(E_1^2+E_2^2+\frac{E_3^2}{f(r)})].\label{det}
\end{equation}

Here the electric field is only turned on the $x^1$-direction
\cite{YS1}, one can take $E_2=E_3=0$ in Eq.(\ref{det}), which
yields
\begin{equation}
det(G_{\mu\nu}+\mathcal{F}_{\mu\nu})=-\frac{r^4}{L^4}f(r)[\frac{r^4}{L^4}-(2\pi
\alpha')^2E^2].\label{det1}
\end{equation}
where we have replaced $E_1$ by $E$ for simplification.

Plugging Eq.(\ref{det1}) into Eq.(\ref{dbi}) and making the
D3-brane located at $r=r_0$, we have
\begin{equation}
S_{DBI}=-T_{D3}\frac{r_0^4}{L^4}\int d^4x
\sqrt{f(r)[1-\frac{(2\pi\alpha^\prime)^2L^4}{r_0^4}E^2]}\label{dbi1}.
\end{equation}

To avoid the action Eq.(\ref{dbi1}) being ill-defined, it is
required that
\begin{equation}
1-\frac{(2\pi\alpha^\prime)^2L^4}{r_0^4}E^2\geq0,
\end{equation}
where we have used the assumption
\begin{equation}
f(r)\geq0.
\end{equation}

So the range of electric field is
\begin{equation}
E\leq\frac{1}{2\pi\alpha^\prime}\frac{r_0^2}{L^2}.
\end{equation}

Finally, the critical field $E_c$ in confining D3-brane background
with chemical potential is obtained, that is,
\begin{equation}
E_c=\frac{1}{2\pi\alpha^\prime}\frac{r_0^2}{L^2}.
\end{equation}

Note that $E_c$ is not affected by the chemical potential. In
other words, the critical field $E_c$ in this case is consistent
with that in confining D3-brane background.

\subsection{total potential} To proceed, we compute the total
potential. For convenience, we introduce a dimensionless parameter
\begin{equation}
\alpha\equiv\frac{E}{E_c}. \label{afa}
\end{equation}

Then, from Eq.(\ref{xx}) and Eq.(\ref{en}), the total potential
$V_{tot}$ can be written as
\begin{eqnarray}
V_{tot}&=&V_{PE+SE}-Ex\nonumber\\&=&2T_Fr_0a\int_1^{1/a}\frac{y^2dy}{\sqrt{(y^4-1)[1-(1+Q^2)(\frac{b}{ay})^4+Q^2(\frac{b}{ay})^6]}}\nonumber\\&-&\frac{2T_F\alpha
r_0}{a}\int_1^{1/a}\frac{dy}{y^2\sqrt{(y^4-1)[1-(1+Q^2)(\frac{b}{ay})^4+Q^2(\frac{b}{ay})^6]}}\label{V}.
\end{eqnarray}

We have checked that the total potential $V_{tot}$ in confining
D3-brane background can be derived from Eq.(\ref{V}) if we neglect
the effect of chemical potential by plugging $Q=0$ in
Eq.(\ref{V}).

\begin{figure}
\centering
\includegraphics[width=8cm]{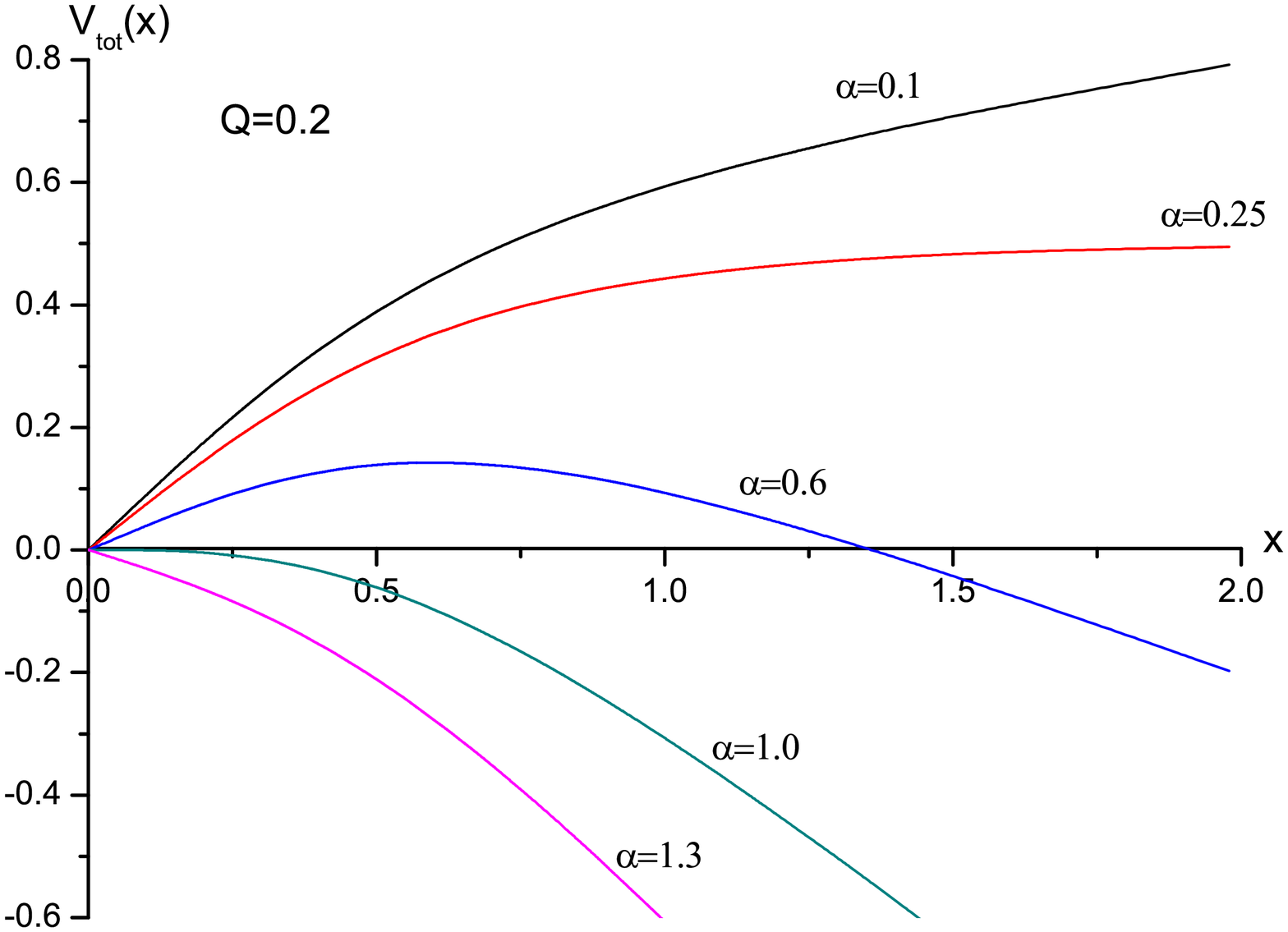}
\includegraphics[width=8cm]{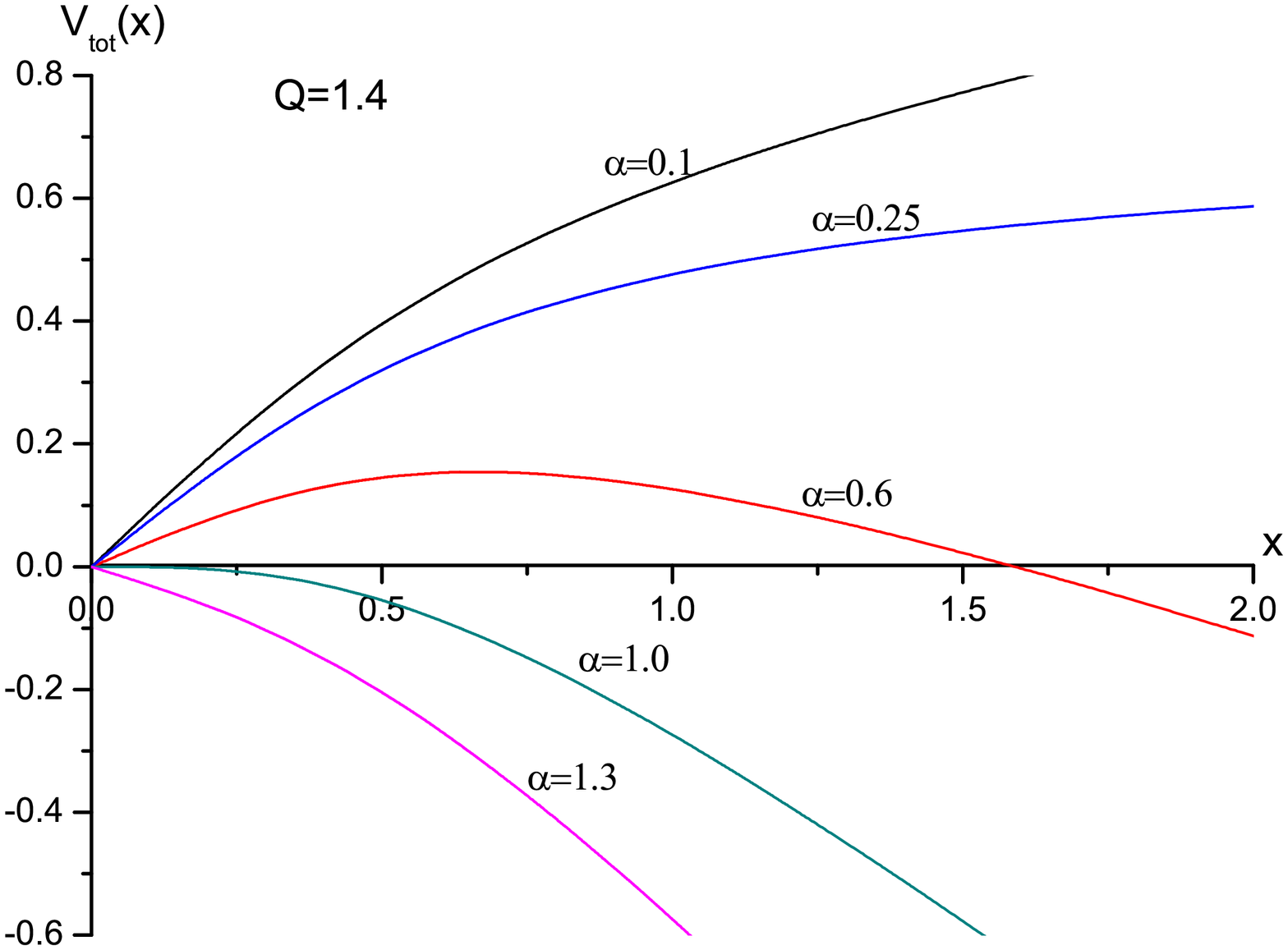}
\caption{The total potential versus distance x with two different
values of Q. Left: $Q=0.2$; Right: $Q=1.4$. In all of the plots
from top to bottom $\alpha=0.1,0.25,0.6,1.0,1.3$. }
\end{figure}

Before discussing the results, let us recall the potential
analysis in confining D3-brane background. There exist two
critical values of the electric field, $E_s=\frac{T_Fr_t^2}{L^2}$
and $E_c=\frac{T_Fr_0^2}{L^2}$. When $E<E_s$, the pair is confined
and no Schwinger effect can occur. When $E_s<E<E_c$, the potential
barrier is present and the pair production is described as a
tunneling process. As E increases, the potential barrier decreases
gradually. At last, it vanishes at $E=E_c$. When $E>E_c$, no
tunneling occurs and the production rate is not exponentially
suppressed any more, the pair production is catastrophic and the
vacuum becomes totally unstable.

We now discuss the results. The total potential can be plotted
versus the separate distance of the test particle pair on the D3-
brane numerically. Here Fig.1 is plotted for a fixed value of b
(fixed temperature) and Q (related to chemical potential) with
different values of $\alpha$. To compare with the case in Ref
\cite{YS1}, we take b=0.5 and
$2L^2/{r_0}={r_0}/(\pi\alpha^\prime)=1$ here. In Fig.1, the left
is plotted for a small value of chemical potential ($Q=0.2$) while
the right is for a larger value of chemical potential ($Q=1.4$).
In all of the plots from top to down
$\alpha=0.1,0.25,0.6,1.0,1.3$. From the figures, we can indeed see
that there exist two critical values of the electric field: one is
related to $\alpha=1$($E=E_c$), the other is at $\alpha=0.25$
($E=E_s=0.25E_c$).

In order to see the effect of chemical potential clearly, we plot
the potential versus x with fixed $\alpha$ and varied Q in Fig.2.
From the left panel in Fig.2, we can see that as Q increases the
height and width of the potential barrier both increase.
Meanwhile, from the right panel in Fig.2, we can see that as Q
increases the height of the potential that becomes flat also
increases. As we know, higher potential barrier makes the produced
pair harder to escape to infinity. So the presence of chemical
potential tends to suppress the Schwinger effect. In addition, as
shown in the right panel in Fig.2, one of the critical values of
the electric field ($E=E_s=0.25E_c$) is not affected by the
chemical potential. This can be also understood by looking at the
value of $E_s$ defined as $E_s=\frac{T_Fr_t^2}{L^2}$.

\begin{figure}
\centering
\includegraphics[width=8cm]{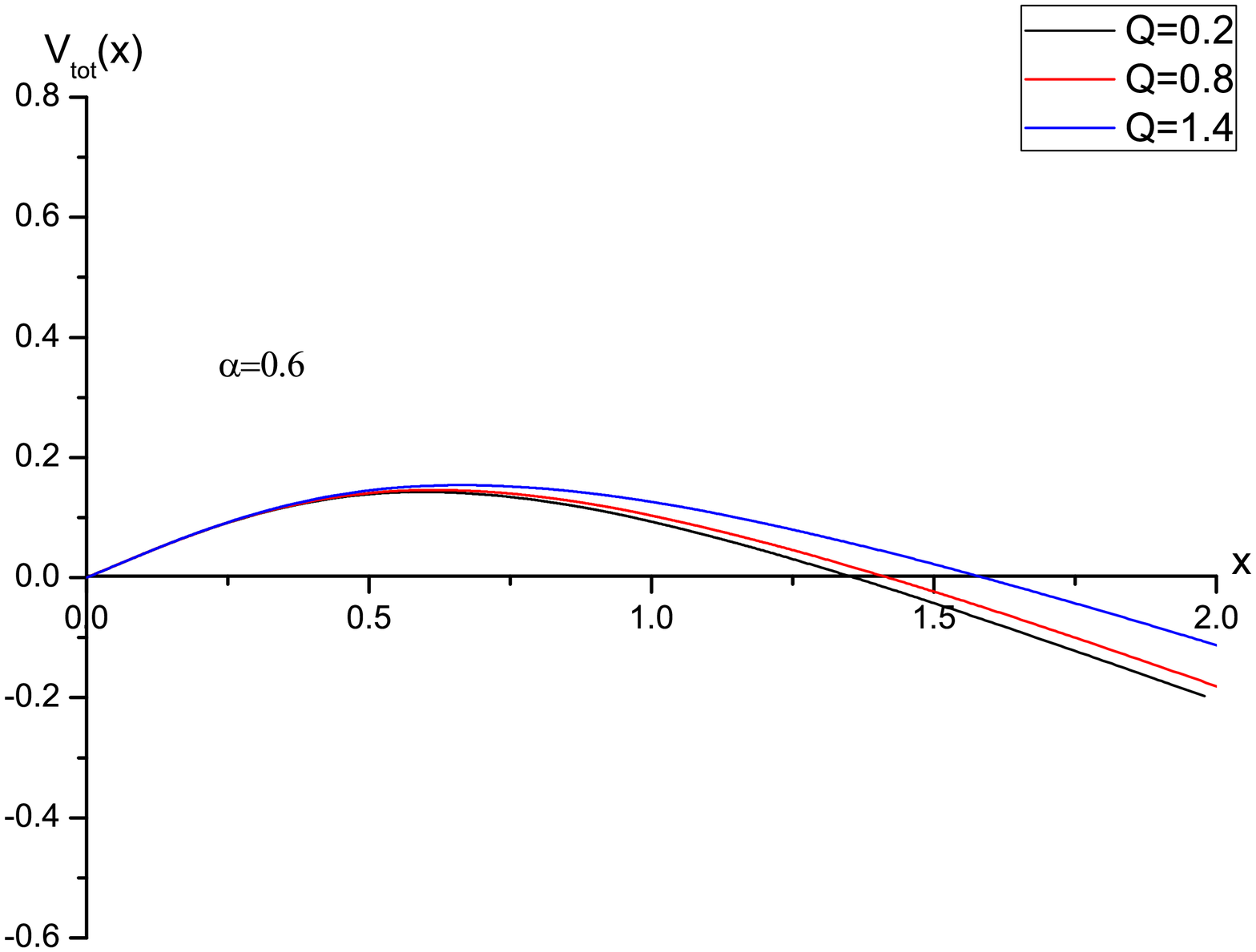}
\includegraphics[width=8cm]{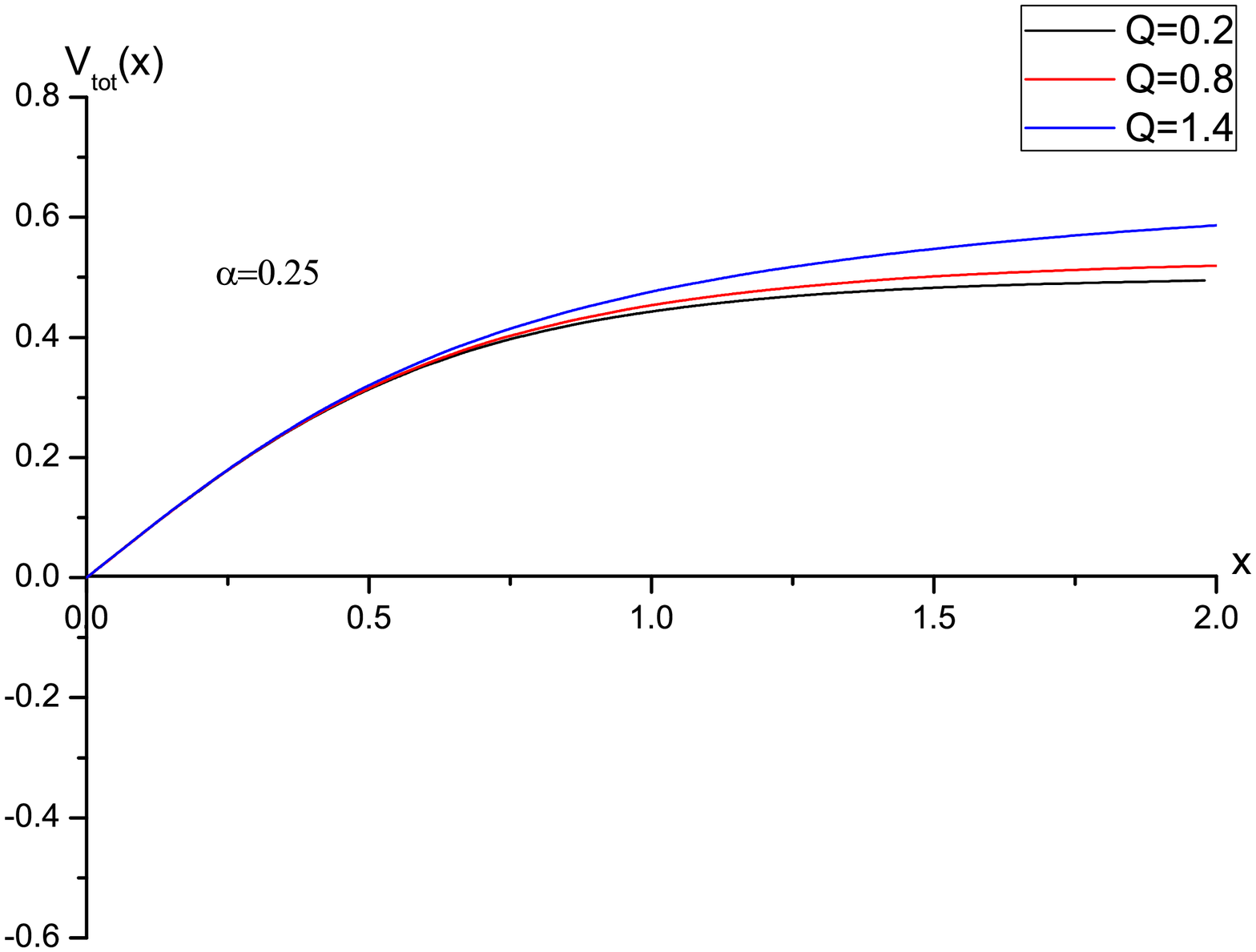}
\caption{The total potential versus distance x with fixed $\alpha$
and varied Q. Left: $\alpha=0.6$; Right: $\alpha=0.25$. In all of
the plots from top to bottom $Q=0.2,0.8,1.4$. }
\end{figure}

\section{Production rate}
In this section, we study the pair production rate. Here we
consider an AdS soliton background \cite{DK}
\begin{eqnarray}
ds^2
=\frac{L^2}{z^2}[-(dx^0)^2+(dx^1)^2+(dx^2)^2+f(z)(dx^3)^2+\frac{dz^2}{f(z)}]+L^2d\Omega^2_5,
\end{eqnarray}
with
\begin{equation}
f(z)=1-(1+Q^2)(\frac{z}{z_t})^4+Q^2(\frac{z}{z_t})^6,
\end{equation}
where the $x^3$-direction is compactified on a circle $S^1$ with
radius $R=\pi z_t$. The AdS boundary is at $z=0$, the geometry is
cut off at $z=z_t$.

The holographic pair production rate is \cite{GW}
\begin{equation}
\Gamma\sim e^{-(S_{NG}+S_{B_2})}.
\end{equation}

To evaluate the pair production rate, one needs to compute the
action or the expectation value of a circular Wilson loop on the
probe brane. In Ref \cite{DK}, the circular Wilson loop is chosen
to lie in the $\tau-\sigma$ plane. As a matter of convenience,
here we use the polar coordinate $(r,\theta)$, as follows from
\cite{SZ}, so the string worldsheet can be parameterized by
$z=z(r)$ in this case. With this ansatz, the induced metric is
given by
\begin{equation}
ds^2=\frac{L^2}{z^2}[(1+\frac{z^{\prime^2}}{f(z)})dr^2+r^2d\theta^2],
\end{equation}
with $z^\prime=\frac{dz}{dr}$.

\begin{figure}
\centering
\includegraphics[width=8cm]{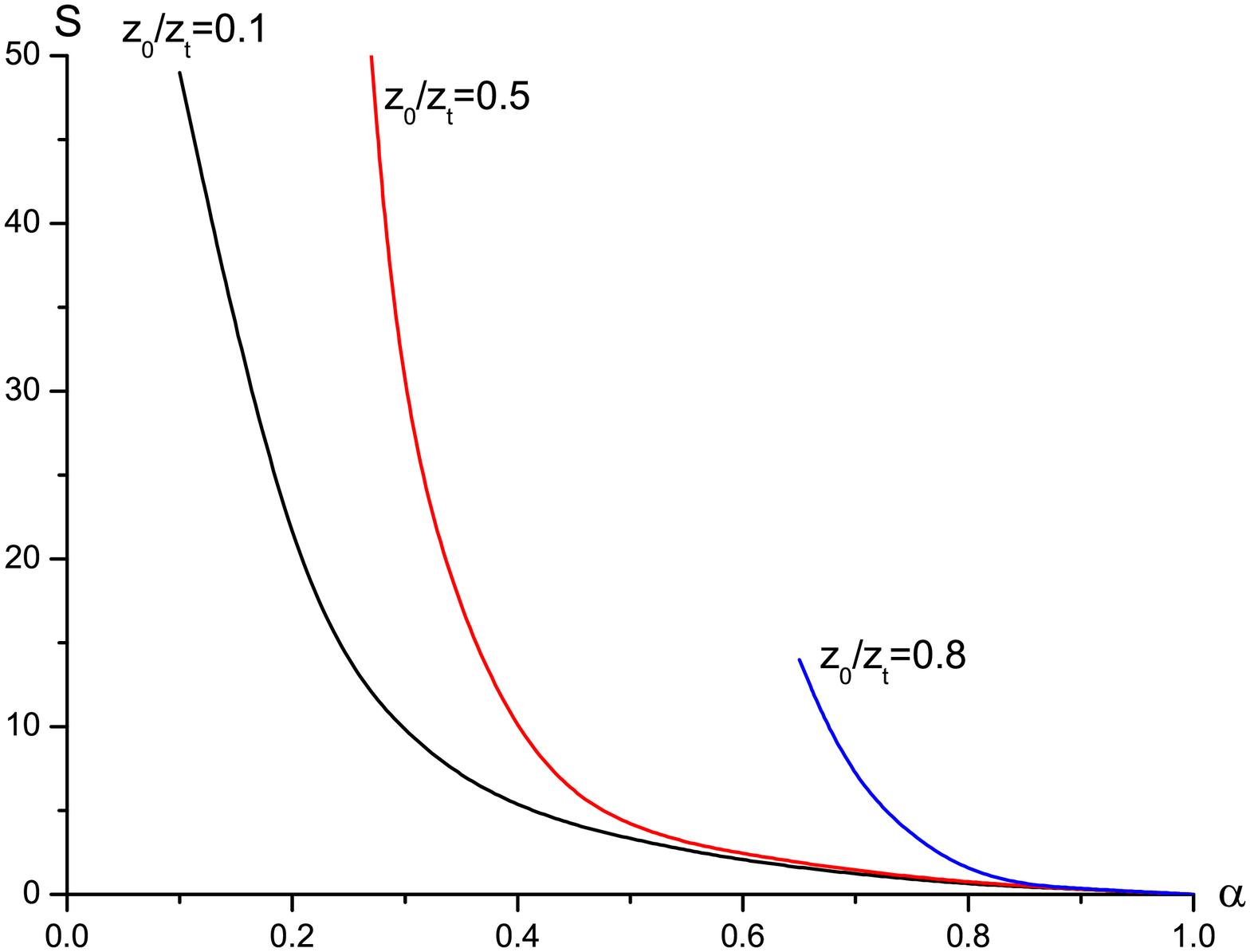}
\includegraphics[width=8cm]{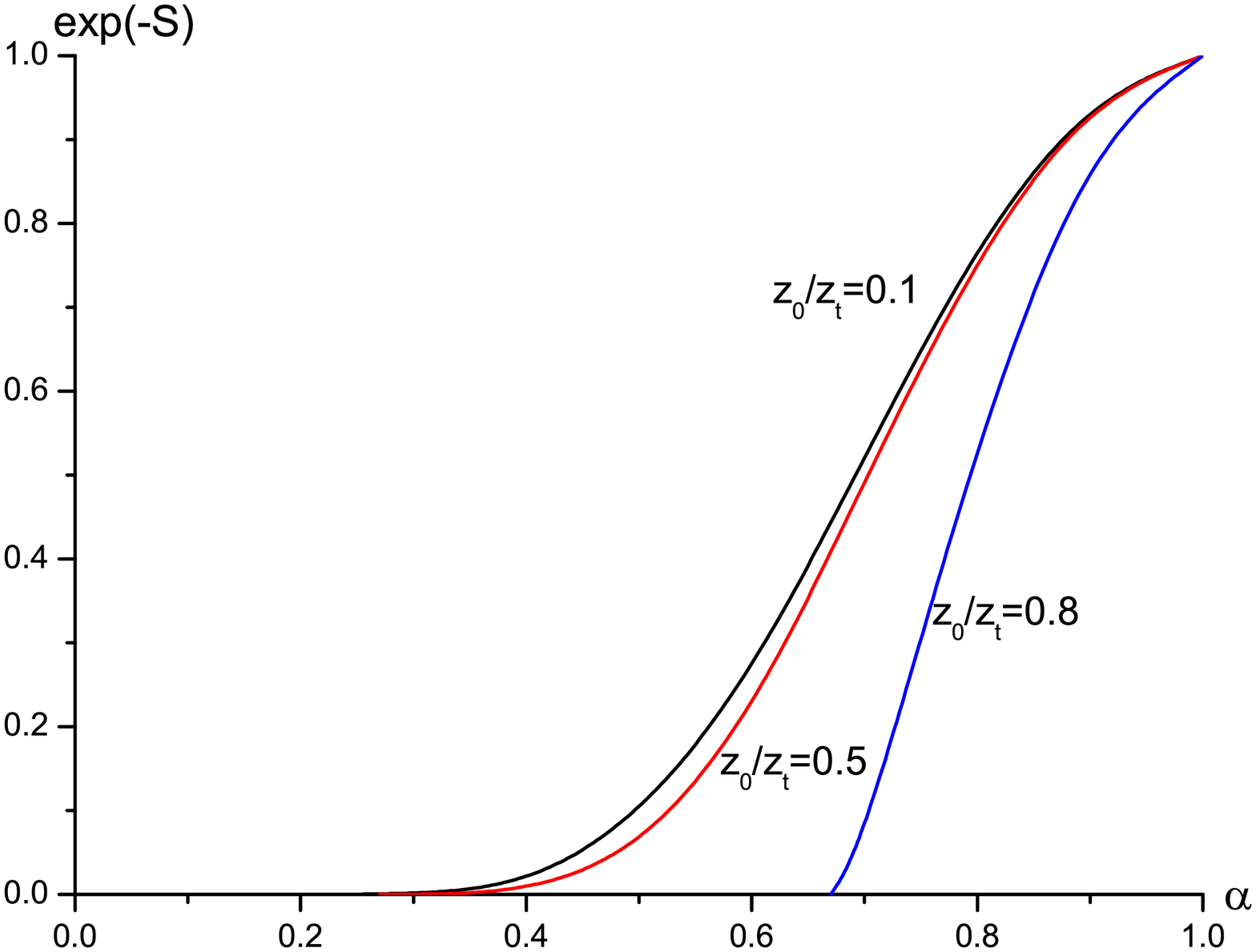}
\caption{The S and $e^{-S}$ versus $\alpha$ with fixed $Q=0.2$. In
all plots from left to right $z_0/z_t=0.1,0.5,0.8$.}
\end{figure}

\begin{figure}
\centering
\includegraphics[width=8cm]{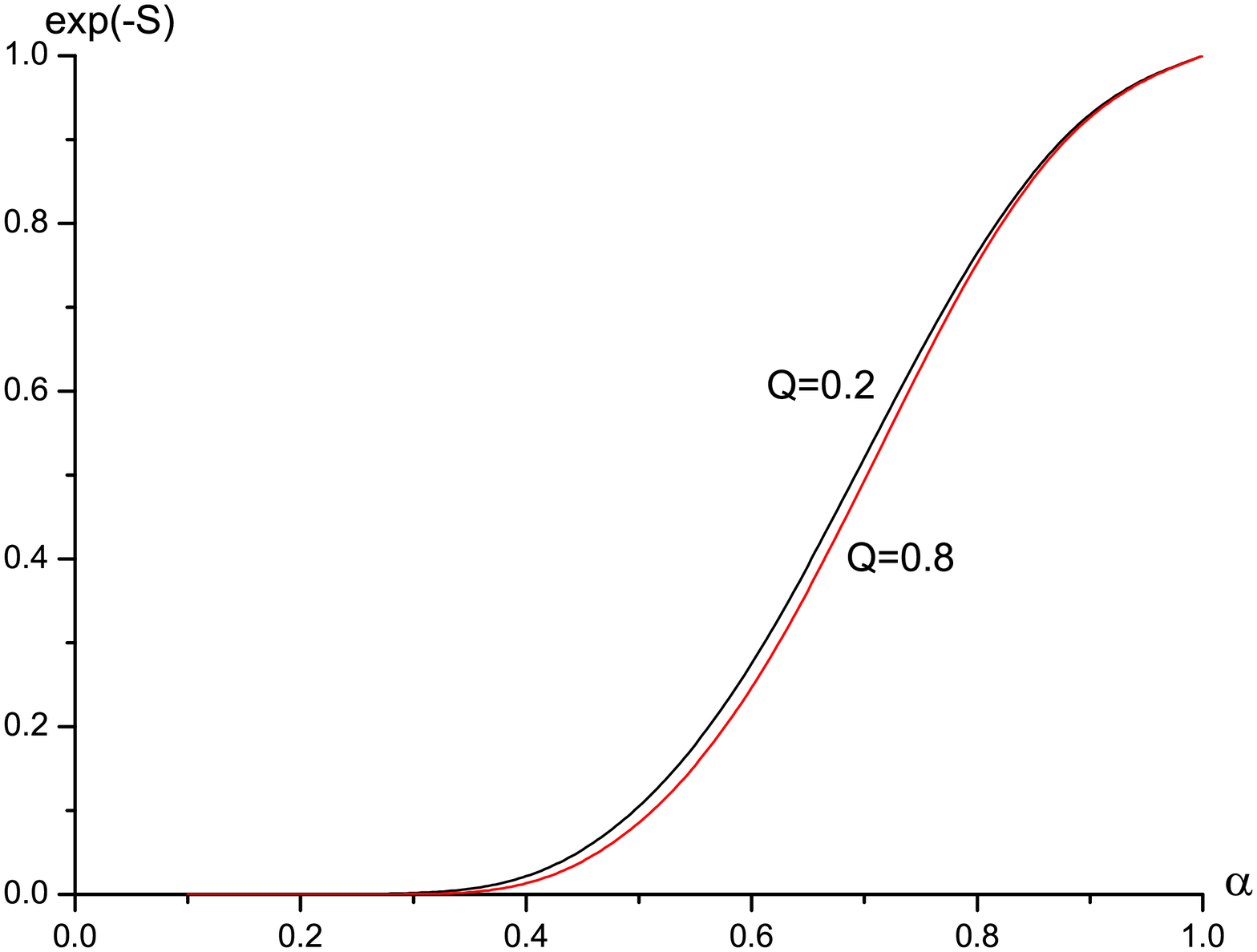}
\includegraphics[width=8cm]{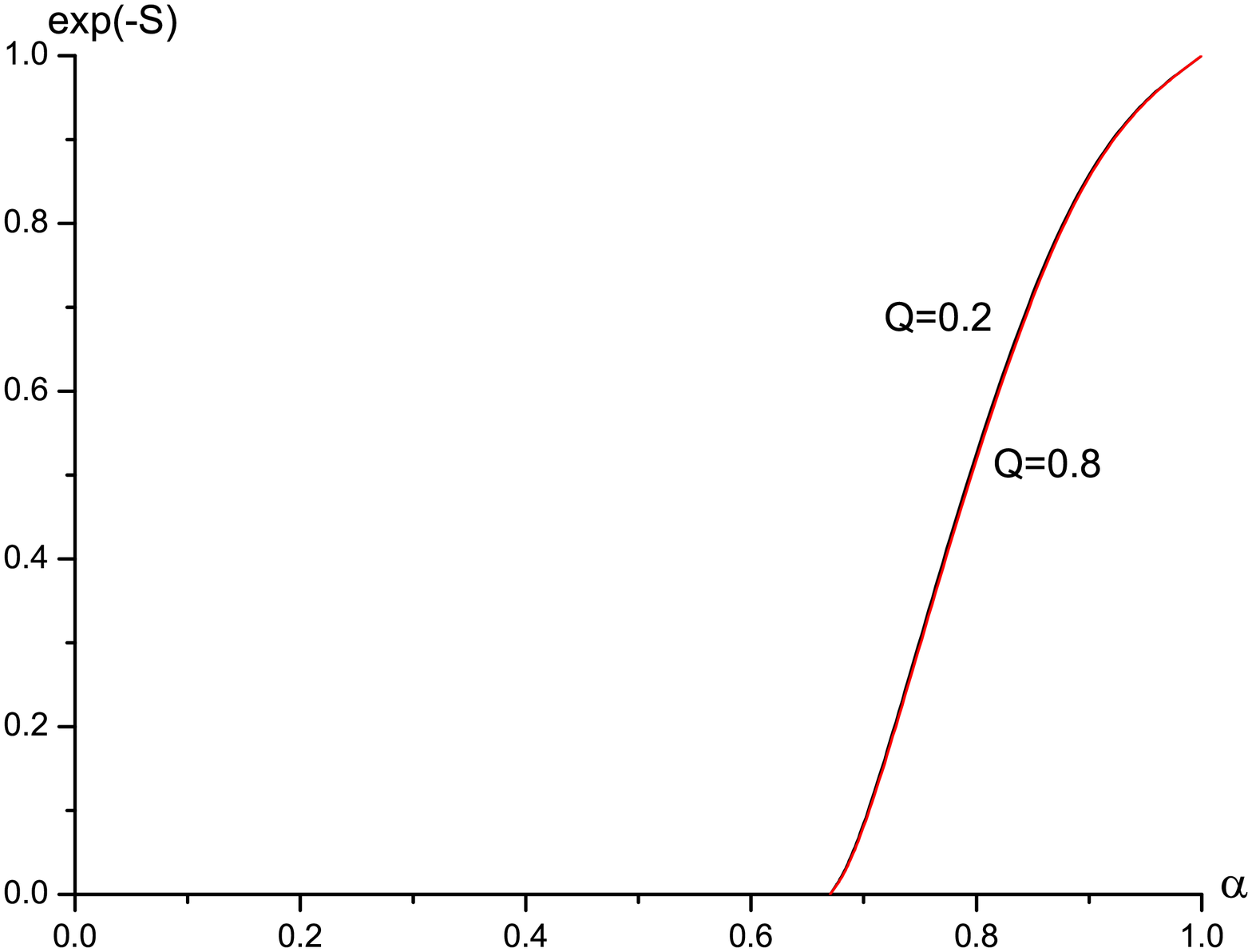}
\caption{The $e^{-S}$ versus $\alpha$ with fixed $z_0/z_t$. Left:
$z_0/z_t=0.1$. Right: $z_0/z_t=0.8$. In all plots from top to
bottom $Q=0.2,0.8$.}
\end{figure}

Then the string action can be expressed as
\begin{equation}
S_{NG}=2\pi
L^2T_F\int_0^xdr\frac{r}{z^2}\sqrt{1+\frac{z^{\prime^2}}{f(z)}},
\end{equation}
\begin{equation}
S_{B_2}=-2\pi T_FB_{01}\int_0^xdr r=-\pi Ex^2,
\end{equation}
where the external electric field is defined as $E\equiv
T_FB_{01}$ with $B_{01}$ the nonvanishing component of $B_2$, x
stands for the radius of the circular Wilson loop. By minimizing
the action, one obtains the equation of motion
\begin{equation}
z^{\prime\prime}(1+r)+\frac{2rz^{\prime^2}}{z}-\frac{rz^{\prime^2}}{2f(z)}\frac{df(z)}{dz}+\frac{2rf(z)}{z}=0.\label{eq1}
\end{equation}

The boundary conditions of the Eq.(\ref{eq1}) are \cite{YS3}
\begin{equation}
z^\prime(z=z_c)=0,\qquad z(r=x)=z_0,\qquad
z^\prime(r=x)=-\sqrt{f(z)(\frac{1}{\alpha^2}-1)}|_{z=z_0},
\end{equation}
where $\alpha$ is defined in Eq.(\ref{afa}).

To proceed further, we need to turn to numerical methods. To
compare with the case in Ref \cite{DK}, we take $2\pi L^2T_F=10$
and set the endpoint of the soliton as $z_t=1$. Then the whole
action is only dependent on $\alpha$, $z_0$ and Q. In Fig.3, we
plot the classical action S and the exponential factor $e^{-S}$
against $\alpha$ with fixed $Q=0.2$. Other cases with different Q
have similar picture. As shown in these figures, the action
diverges around a certain value which leads to a vanishing
$e^{-S}$. We know from the above section that when
$\alpha<\alpha_s$, with
$\alpha_s\equiv\frac{E_s}{E_c}=\frac{z_0^2}{z_t^2}$, the Schwinger
effect does not occur. Thus, this agreement supports that the
numerical results of the production rate are consistent with the
potential analysis.

To see the effect of the chemical potential on Schwinger effect,
we plot $e^{-S}$ versus $\alpha$ at fixed $z_0/z_t$ with two
different values of Q in Fig.4. From the plots, we can see that
$e^{-S}$ decreases as Q increases. This means that the chemical
potential suppresses the pair production rate, also agrees with
the previous potential analysis. Moreover, as $z_0/z_t$ increases,
the effect of the chemical potential becomes weaker, this
phenomenon can be found in the right panel in Fig.4, where the two
curves with different Q are nearly overlapped.

\section{conclusion and discussion}

Schwinger effect in confining gauge theories may be an important
ingredient in looking for new aspects of QCD in the presence of
strong external fields. In this paper, we have studied the effect
of chemical potential on the holographic Schwinger effect by
considering a confining D3-brane background with charge. The
potential of a test particle pair was obtained by calculating the
Nambu-Goto action of string attaching the rectangular Wilson loop
at the probe D3-brane. The production rate for various cases was
evaluated numerically. The effect of the chemical potential on the
production rate was qualitatively shown. From the results, we can
indeed see that there exist two critical values of the electric
field. In addition, both the potential analysis and the numerical
results of the production rate suggest that the presence of
chemical potential tend to suppress the Schwinger effect.
Interestingly, the instanton effects on the Schwinger effect has
been studied in Ref \cite{MD}. It is found that the presence of
instantons also suppresses the pair production effect at finite
temperature.

\section{Acknowledgments}

This research is partly supported by the Ministry of Science and
Technology of China (MSTC) under the ¡°973¡± Project no.
2015CB856904(4). Zi-qiang Zhang is supported by the NSFC under
Grant no. 11547204. Gang Chen is supported by the NSFC under Grant
no. 11475149. De-fu Hou is partly supported by NSFC under Grant
nos. 11375070 and 11221504.


\end{document}